\documentclass[aps,prl,showpacs,twocolumn,lineno,groupedaddress]{revtex4}

\usepackage{graphicx}
\usepackage{epsfig}
\usepackage{dcolumn}
\usepackage{bm}
\usepackage{amssymb}   
\usepackage[usenames]{color} 

\newcommand{\beq}{\begin{equation}}
\newcommand{\eeq}{\end{equation}}
\newcommand{\barr}{\begin{eqnarray}}
\newcommand{\earr}{\end{eqnarray}}

\def\bq{\begin{quote}}
\def\eq{\end{quote}}

\def\spose#1{\hbox to 0pt{#1\hss}}
\def\lsim{\mathrel{\spose{\lower 3pt\hbox{$\mathchar"218$}}
 \raise 2.0pt\hbox{$\mathchar"13C$}}}
\def\gsim{\mathrel{\spose{\lower 3pt\hbox{$\mathchar"218$}}
 \raise 2.0pt\hbox{$\mathchar"13E$}}}

\def\bsbar{${\overline{B}_s^0}$}
\def\bs{${B_s^0}$}

\def\bsdec{${B_s^0 \rightarrow J/\psi \phi}$}

\def\bddec{${B_d^0 \rightarrow J/\psi K^*}$}

\def\ADD{$|A_{0}(0)|^{2}-|A_{||}(0)|^{2}$}

\def\D0{D\O }

\def\GeVp{ {\ifmmode \;{{\mbox{\mathrm GeV}} / {\mbox\mathrm c}} \else
${{\mbox{\mathrm GeV}} / {\mbox\mathrm c}}$ \fi }}
\def\MeVp{ {\ifmmode \;{{\mbox{\mathrm MeV}} / {\mbox\mathrm c}} \else
${{\mbox{\mathrm MeV}} / {\mbox\mathrm c}}$ \fi }}
\def\MeV{ {\ifmmode \;{{\mbox{\mathrm MeV}} / {\mbox\mathrm c}^2} \else
${{\mbox{\mathrm MeV}} / {\mbox\mathrm c}^2}$ \fi }}
\def\GeV{ {\ifmmode \;{{\mbox{\mathrm GeV}} / {\mbox\mathrm c}^2} \else
${{\mbox{\mathrm GeV}} / {\mbox\mathrm c}^2}$ \fi }}

\providecommand{\tabularnewline}{\\}

\begin{document}


%

\hspace{5.2in}
\mbox{FERMILAB-PUB-08-033-E}

\title{
Measurement of $\boldmath {B_s^0}$ mixing parameters from the flavor-tagged decay 
{\boldmath \bsdec}  
}
\date{February 15, 2008}

%
\author{V.M.~Abazov$^{36}$}
\author{B.~Abbott$^{75}$}
\author{M.~Abolins$^{65}$}
\author{B.S.~Acharya$^{29}$}
\author{M.~Adams$^{51}$}
\author{T.~Adams$^{49}$}
\author{E.~Aguilo$^{6}$}
\author{S.H.~Ahn$^{31}$}
\author{M.~Ahsan$^{59}$}
\author{G.D.~Alexeev$^{36}$}
% ...

\affiliation{\vspace{0.1 in}(The D\O\ Collaboration)\vspace{0.1 in}}
% ...
\affiliation{$^{6}$University of Alberta, Edmonton, Alberta, Canada,
                Simon Fraser University, Burnaby, British Columbia, Canada,
                York University, Toronto, Ontario, Canada, and
                McGill University, Montreal, Quebec, Canada}
% ...
\affiliation{$^{29}$Tata Institute of Fundamental Research, Mumbai, India}
% ...
\affiliation{$^{31}$Korea Detector Laboratory, Korea University, Seoul, Korea}
% ...
\affiliation{$^{36}$Joint Institute for Nuclear Research, Dubna, Russia}
% ...
\affiliation{$^{49}$Florida State University, Tallahassee, Florida 32306, USA}
% ...
\affiliation{$^{51}$University of Illinois at Chicago,
                Chicago, Illinois 60607, USA}
% ...
\affiliation{$^{59}$Kansas State University, Manhattan, Kansas 66506, USA}
% ...
\affiliation{$^{65}$Michigan State University,
                East Lansing, Michigan 48824, USA}
% ...
\affiliation{$^{75}$University of Oklahoma, Norman, Oklahoma 73019, USA}
% ...

           
\begin{abstract}

From an analysis of the flavor-tagged decay \bsdec\
we obtain the width difference between the $B_s^0$ light  and  
heavy  mass eigenstates,
$\Delta \Gamma_s \equiv \Gamma_L - \Gamma_H
= 0.19 \pm 0.07 {\rm (stat)}\thinspace ^{+0.02}_{-0.01} {\rm (syst)}$ ps$^{-1}$, and 
the $CP$-violating phase,  
$\phi_{s} =-0.57 ^{+0.24}_{-0.30} {\rm (stat)}\thinspace ^{+0.07}_{-0.02} {\rm (syst)}$. 
The allowed 90\% C.L. intervals of $\Delta \Gamma_s$ and  $\phi_s$ 
are $0.06 <\Delta \Gamma_s  <0.30$  ps$^{-1}$ and
 $-1.20 <\phi_s < 0.06$, respectively.
The data sample corresponds to an integrated luminosity of 2.8 fb$^{-1}$
accumulated with the D0 detector at the Fermilab Tevatron collider.

\end{abstract}


\maketitle

\newpage

In the standard model (SM), the light ($L$) and heavy ($H$) mass eigenstates 
of the mixed $B_s^0$ system  are expected
to have  sizeable 
mass and decay width differences: $\Delta M_s \equiv M_H - M_L$ and
$\Delta \Gamma_s \equiv \Gamma_L - \Gamma_H$. 
The two mass eigenstates are expected to be almost pure $CP$ 
eigenstates.  
The $CP$-violating mixing phase
that appears in $b \rightarrow c \overline c  s$ decays
is predicted~\cite{LN2006}
to be $\phi_s = -2\beta_s = 2\arg[-V_{tb}V^*_{ts}/V_{cb}V^*_{cs}]
 = -0.04 \pm 0.01$, where
$V_{ij}$ are elements of the  Cabibbo-Kobayashi-Maskawa 
quark-mixing matrix~\cite{ckm}. 
New phenomena may alter the phase to $\phi_s \equiv -2\beta_s +\phi_s^\Delta$.

In Ref.~\cite{prl07}, we presented an analysis 
of  the  decay chain \bsdec, 
$J/\psi \rightarrow \mu ^+ \mu ^-$,
$\phi \rightarrow K^+ K^-$ based on  1.1 fb$^{-1}$ 
of data collected  with the  D0 detector~\cite{run2det}
at the Fermilab Tevatron collider.
In that analysis we measured $\Delta \Gamma_s$ and
the average lifetime of the \bs\ system, 
$\overline \tau_s =1/\overline \Gamma_s$, where
$\overline\Gamma_s \equiv(\Gamma_H+\Gamma_L)/2$.
The $CP$-violating phase $\phi_s$ was also extracted 
for the first time.
The measurement  correlated two solutions for $\phi_s$ with 
two corresponding solutions for $\Delta \Gamma_s$.
Improved precision was obtained by refitting the results
using additional experimental constraints~\cite{combo}.
 Here we present new D0 results of an analysis that
includes information on the
$B_s^0$ flavor at production time. 
Adding this information resolves the sign ambiguity 
on $\phi_s$ for a given $\Delta \Gamma_s$
and improves the precision of the measurement. 
The analysis is based on an increased data 
set, collected between October 2002 and June 2007, and 
corresponding to 
 an integrated luminosity of 2.8  fb$^{-1}$.

\begin{figure}[htb]
\begin{center}\includegraphics[%
  width=6.9cm,
  keepaspectratio,
  trim=0 40 0 0
  ]{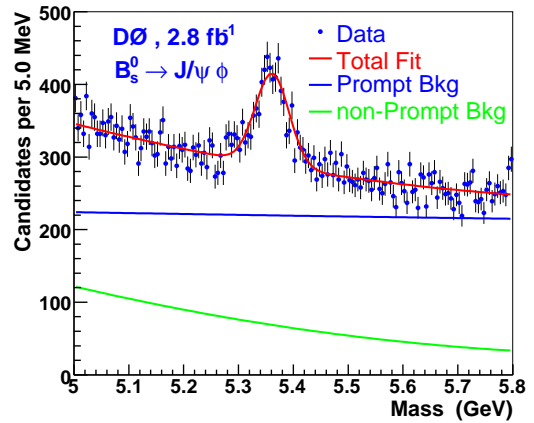}\end{center}

\caption{\label{fig:mass}
The invariant mass distribution of the ($J/\psi,\phi$) system for 
 \bs\ candidates. The curves are projections of the maximum likelihood fit (see text). 
}
\end{figure}

We reconstruct the decay chain \bsdec, $J/\psi \rightarrow \mu ^+ \mu ^-$,
$\phi \rightarrow K^+ K^-$ from  candidate  ($J/\psi,\phi$) pairs
 consistent with coming from a common vertex and  having 
an invariant mass in the range 5.0 -- 5.8 GeV. 
The event selection follows that in Ref.~\cite{prl07}.
The invariant mass distribution of the  48047 candidates  is
shown in Fig.~\ref{fig:mass}. The curves are   
projections of the maximum likelihood fit, described below.
The fit assigns  1967$\pm$65 (stat) events to the \bs\ decay.
The flavor of the initial state of the $B_s^0$ candidate is determined
by exploiting the properties of particles produced
by the other $b$ hadron  (``opposite-side 
tagging'') and the properties of particles accompanying the 
$B_s^0$ meson (``same-side tagging'').
The variables used to construct the opposite-side 
tagging are described  in Ref.~\cite{bflavor}. 
The only difference to the description in 
Ref.~\cite{bflavor}  is that the events 
that  do not contain
either the opposite lepton or the secondary vertex, and that were
not used for the flavor tagging before, are now tagged with the
event-charge variable defined in Ref.~\cite{bflavor}.

Same-side
tagging is based on the sign of an associated charged kaon
formed in the hadronization process.
A  $B_s^0~(\bar b s)$  meson is expected to be accompanied
by a strange meson, e.g.  $K^+ ~(u \bar s)$  meson that can
be used for flavor tagging. Such a configuration 
is formed when 
the initial $\bar b$ antiquark picks up an  $s$ quark from a virtual 
$s \bar s$ pair and the $\bar s$ antiquark becomes a
constituent of an accompanying $K^+$  meson.  
Candidates for the associated kaon are
all charged tracks with transverse momentum $p_T > 500$ MeV that are
not used in the $B_s^0$ reconstruction.
We define the quantity
$\Delta R = \sqrt{(\Delta \phi)^2 + (\Delta \eta)^2} $, where
$\Delta \phi$ ($\Delta \eta$) is the distance in the azimuthal angle
(pseudorapidity) between the given
track and the $B_s$ meson, and
select the track with the minimum value of $\Delta R$.
The corresponding discriminating
variable for the flavor tagging is defined as the
product of the particle charge and $\Delta R$.
 Another discriminating
variable is $Q_{\rm jet}$, the
$p_T$-weighted   average of all track charges $q_i$
within  the cone
$\cos[ \angle (\vec {p}, {\vec p}_B )] > 0.8$ around the $B$ meson:
$Q_{\rm jet} = [\sum_i q^i (p_T^i)^{0.6}] / \sum_i (p_T^i)^{0.6}$.

The discriminating variables of both the same-side and opposite-side
tagging are combined  using the likelihood-ratio method described 
in Ref.~\cite{bflavor}. A tag is defined for 99.7\% of events.
The performance of the combined tagging is taken from a Monte Carlo (MC) simulation
of the \bsdec\ process and
is verified with the  $B^\pm \to J/\psi K^\pm$ process
for which we find the simulated tagging to be in agreement with data.
The effective tagging power, as defined in
Ref~\cite{bflavor}, is  ${\cal{P}} = (4.68 \pm 0.54)\%$. It is a
significant improvement over the performance of the
opposite-side tagging alone,
${\cal{P}} = (2.48 \pm 0.22)\%$~\cite{bflavor}.
The purity of the flavor tag as a function of an over-all flavor
discriminant is determined and parametrized, and
the related probability ${\rm P}(B_s)$ of having  
a pure state   $B_s^0$  at $t=0$   is
used event-by-event in the fit described below.

We perform an unbinned maximum likelihood fit to the proper decay time, 
three decay angles characterizing the final state, and mass of the $B_s^0$ candidate. 
The likelihood function ${\cal L}$ is given by:
\begin{eqnarray}
{\cal L} & = & \prod^{N}_{i=1}[ f_{\text {sig}}{\cal F}^i_{\text {sig}} + 
(1-f_{\text {sig}}){\cal F}^i_{\text {bck}}],
\end{eqnarray}
where $N$  is the total number of events, 
and $f_{\text {sig}}$ is the fraction of signal in the sample.
The function  ${\cal F}^i_{\text {sig}}$  describes the distribution of the
signal in  mass, proper decay time, and the decay angles.
For the signal mass distribution, we use
a  Gaussian function with  free mean and width.
The proper decay time distribution of the $L$ or $H$ component
of the signal is parametrized  by an
exponential convoluted with a Gaussian function.
The  width of the Gaussian is
taken from  the  event-by-event 
estimate of  the $ct$  uncertainty   $\sigma(ct)$,  scaled by an  overall 
calibration factor  determined from
the fit  to the prompt component of the background. 
${\cal F}^i_{\text {bck}}$ is the product of the background mass, proper decay time, 
and angular  probability density functions.
Background is divided into two categories. 
``Prompt'' background is due to directly 
produced $J/\psi$ mesons accompanied by random tracks arising from 
hadronization.  This background is distinguished from ``non-prompt'' 
background, where the $J/\psi$ meson is a product of a $B$-hadron decay 
while the tracks forming the $\phi$ candidate emanate from a multibody 
decay of a $B$ hadron or from hadronization.

The decay amplitude of the $B_s^0$ and $\overline B_s^0$ mesons 
is decomposed into three independent
components corresponding to linear polarization states of 
the vector mesons  $J/\psi$ and  $\phi$,
which are either 
longitudinal (0) or transverse to their direction of motion,
and parallel ($\parallel$) or perpendicular ($\perp$) to each other.
The time evolution of the angular distribution of the decay products,
expressed in terms of 
the magnitudes  $|A_0|$, $|A_\parallel|$, and  $|A_\perp|$,
 and two relative strong phases   
$\delta _1 = -\delta_{||} + \delta_\perp$ and
 $\delta _2 = -\delta_0 + \delta_\perp$ of the amplitudes,
is given in Ref.~\cite{DFN2001}: 

\begin{eqnarray}
\lefteqn{ \frac {d^4 \Gamma } {dt d\cos\theta d\varphi d\cos\psi} \propto } \nonumber \\
& & 2 \cos^2\psi (1 - \sin^2 \theta\cos^2 \varphi) |A_0(t)|^2 \nonumber \\
& &+ \sin^2 \psi (1 - \sin^2 \theta \sin^2 \varphi)    {|A_\parallel(t)|^2 } \nonumber \\
& &+ \sin^2 \psi \sin^2 \theta     {  |A_\perp(t)|^2  } \nonumber \\
& &+ (1/\sqrt{2}) \sin 2 \psi \sin^2\theta \sin 2 \varphi {{\mathrm {Re}} (A_0^{*}(t) A_\parallel(t))} 
\nonumber \\
& &+ \ \ (1/\sqrt{2}) \sin2\psi \sin2\theta\cos\varphi   {{\mathrm {Im}}(A_0^{*}(t) A_\perp(t))} \nonumber \\
& &- \ \sin^2\psi \sin2\theta\sin\varphi   { {\mathrm  {Im}}(A_\parallel^{*}(t) A_\perp(t)). }
\end{eqnarray}

Polarization amplitudes for $B_s^0$ (upper sign) and 
$\overline {B}_s^0$ (lower sign) are given by the following equations:

\begin{eqnarray*}
|A_{0,\parallel}(t)|^2 & = & |A_{0,\parallel}(0)|^2 \left[{  {\cal{T_{+}}} } \pm e^{-\overline{\Gamma}t} \sin\phi_{s} \ \sin(\Delta M_{s} t) \right],  \\ 
 {  |A_\perp(t)|^2 } &  = & |A_{\perp}(0)|^2 \left[{  {\cal{T_{-}}} }  \mp  e^{-\overline{\Gamma}t}  \sin\phi_{s} \   \sin(\Delta M_{s} t)\right], 
\end{eqnarray*}
\begin{eqnarray*}
\lefteqn{{\mathrm {Re}} (A_0^{*}(t) A_\parallel(t)) = |A_0(0)||A_{\|}(0)|\cos(\delta_2 - \delta_1)}  \\
& & \times \left[{ {\cal{T_{+}}} }  \pm  e^{-\overline{\Gamma}t} \sin\phi_{s} \ \sin(\Delta M_{s} t) \right],  
\end{eqnarray*}
\begin{eqnarray*}
\lefteqn{{\mathrm {Im}}(A_0^{*}(t) A_\perp(t))  =  |A_0(0)||A_{\perp}(0)|| } \\
& & \times [e^{-\overline{\Gamma}t} ( {\  \pm} \sin\delta_2 \cos(\Delta M_{s} t) {\  \mp} \cos\delta_2 \sin(\Delta M_{s} t)\cos\phi_{s} )   - \\
& & (1/2) (e^{-\Gamma_H t}-e^{-\Gamma_Lt}) \sin\phi_{s} \ \cos\delta_2],
\end{eqnarray*}
\begin{eqnarray*}
\lefteqn{ { {\mathrm  {Im}}(A_\parallel^{*}(t) A_\perp(t)) } = |A_{\|}(0)||A_{\perp}(0)|} \\
& &\times [  \ e^{-\overline{\Gamma}t} ( {\  \pm} \sin\delta_1 \cos(\Delta M_{s} t) \mp \cos\delta_1 \sin(\Delta M_{s} t)\cos\phi_{s} ) \\
& &  - (1/2) (e^{-\Gamma_H t}-e^{-\Gamma_Lt}) \sin\phi_{s} \ \cos\delta_1],
\end{eqnarray*}
where 
${\cal{T_{\pm}}} = (1/2) \left[{ (1 \pm \cos\phi_{s})e^{-\Gamma_{L}t} +  (1 \mp \cos\phi_{s})e^{-\Gamma_{H}t}} \right].$ 
For a given event, the decay rate is the sum of the 
$B_s^0$ and $\overline B_s^0$ rates weighted  by
${\rm P}(B_s)$ and $1-{\rm P}(B_s)$, respectively, and by the 
detector acceptance.

In the coordinate system of the $J/\psi$ rest frame 
 (where the $\phi$ meson moves in the $x$ direction,
 the $z$  axis is perpendicular to 
the decay plane of $\phi \to K^+ K^-$, and $p_y(K^+)\geq 0$),
the transversity polar and azimuthal angles 
$(\theta, \varphi)$ describe the
direction of the $\mu^+$, and $\psi$ is 
the angle between   $\vec p(K^+)$ and  $-\vec{p}(J/\psi)$ 
 in the $\phi$ rest frame.

We model the acceptance and resolution of the three angles by fits 
using polynomial functions, with parameters determined using 
MC simulations.
Events generated uniformly in the three-angle space
were processed through the standard  {\sc GEANT}-based~\cite{geant} 
simulation of the D0 detector, and reconstructed and selected 
as real data.
Simulated events 
were reweighted to match the kinematic distributions observed in
the data.

The proper decay time distribution shape  of the  background 
is described as a sum of
a prompt component, modeled as 
a Gaussian function centered at zero, and a non-prompt component.
The non-prompt component  is modeled as a superposition of one 
exponential for $t<0$ 
and two exponentials for $t>0$, with free slopes and normalizations.
The distributions of the  backgrounds in mass,
$\cos\theta$, $\varphi$, and  $\cos\psi$ 
are parametrized by low-order polynomials.
We also allow for a background term analogous to the interference term
of the $A_0$ and $A_\parallel$  waves,
with one free coefficient. For each of the 
above background functions  we use two separate sets of parameters 
for  the prompt and non-prompt  components.

The high degree of correlation between $\Delta M_s$,  $\phi_s$,
and the two $CP$-conserving strong phases $\delta_1$ and $\delta_2$
makes it difficult 
to obtain stable fits when all of them are allowed to vary freely.
 In the following, 
we fix  $\Delta M_s$ to  $17.77 \pm 0.12$ ps$^{-1}$,
as measured in Ref.~\cite{dms}.
The phases analogous to $\delta_i$ have been measured 
for the decay \bddec\  at the $B$ factories.
We allow the phases
$\delta_i$ to vary around the
the world-average values~\cite{wadelta12} 
for the \bddec\ decay, $\delta_1 = -0.46$ and $\delta_2 = 2.92$, under 
a Gaussian constraint. The width  of the Gaussian, chosen to be 
$\pi/5$, allows for 
some degree of violation of the $SU(3)$ symmetry relating the two 
decay processes, 
while still effectively constraining the signs of 
$\cos\delta_i$ to agree with those of Ref.~\cite{wadelta12}. 
The mirror solution with $\cos\delta_1<0$ is  disfavored
on theoretical~\cite{suzuki} and experimental~\cite{babar} grounds.

\begin{table}[h!tb]
\caption {Summary of the likelihood fit results for three cases:
free $\phi_s$, 
$\phi_s$ constrained to  the SM  value, and  
$\Delta\Gamma_s$ constrained by the expected relation $\Delta\Gamma_s^{SM}\cdot|\cos(\phi_s)|$. }
\renewcommand{\arraystretch}{1.2}
\begin{center}
\begin{tabular}{c|c|c|c}
\hline
\hline
           & free $\phi_s$     & $\phi_s \equiv \phi_s^{SM}$  & $\Delta\Gamma_s^{th}$  \tabularnewline 
\hline
$\overline{\tau}_s$ (ps)     & 1.52$\pm$0.06  & 1.53$\pm$0.06 & 1.49$\pm$0.05  \tabularnewline
$\Delta\Gamma_s$ (ps$^{-1}$) & 0.19$\pm$0.07  & 0.14$\pm$0.07 & $0.083 \pm 0.018$ \tabularnewline
$A_{\perp}(0)$               & 0.41$\pm$0.04  & 0.44$\pm$0.04 & 0.45$\pm$0.03  \tabularnewline
\ADD                         & 0.34$\pm$0.05  & 0.35$\pm$0.04 & 0.33$\pm$0.04 \tabularnewline
$\delta_1$                   &$-0.52\pm$0.42  &$-0.48\pm$0.45 &$-0.47\pm$0.42  \tabularnewline
$\delta_2$                   & 3.17$\pm$0.39  & 3.19$\pm$0.43 &3.21$\pm$0.40 \tabularnewline
$\phi_s$                     &$-0.57^{+0.24}_{-0.30}$  & $\equiv -0.04$ & $-0.46 \pm 0.28$\tabularnewline
$\Delta M_s$ (ps$^{-1}$)     & $\equiv$ 17.77    & $\equiv$ 17.77 &$\equiv$ 17.77 \tabularnewline
\hline
\hline
\end{tabular}
\label{tabmainres}
\end{center}
\end{table}

\begin{figure}[h!tb]
\begin{center}
\psfig{file=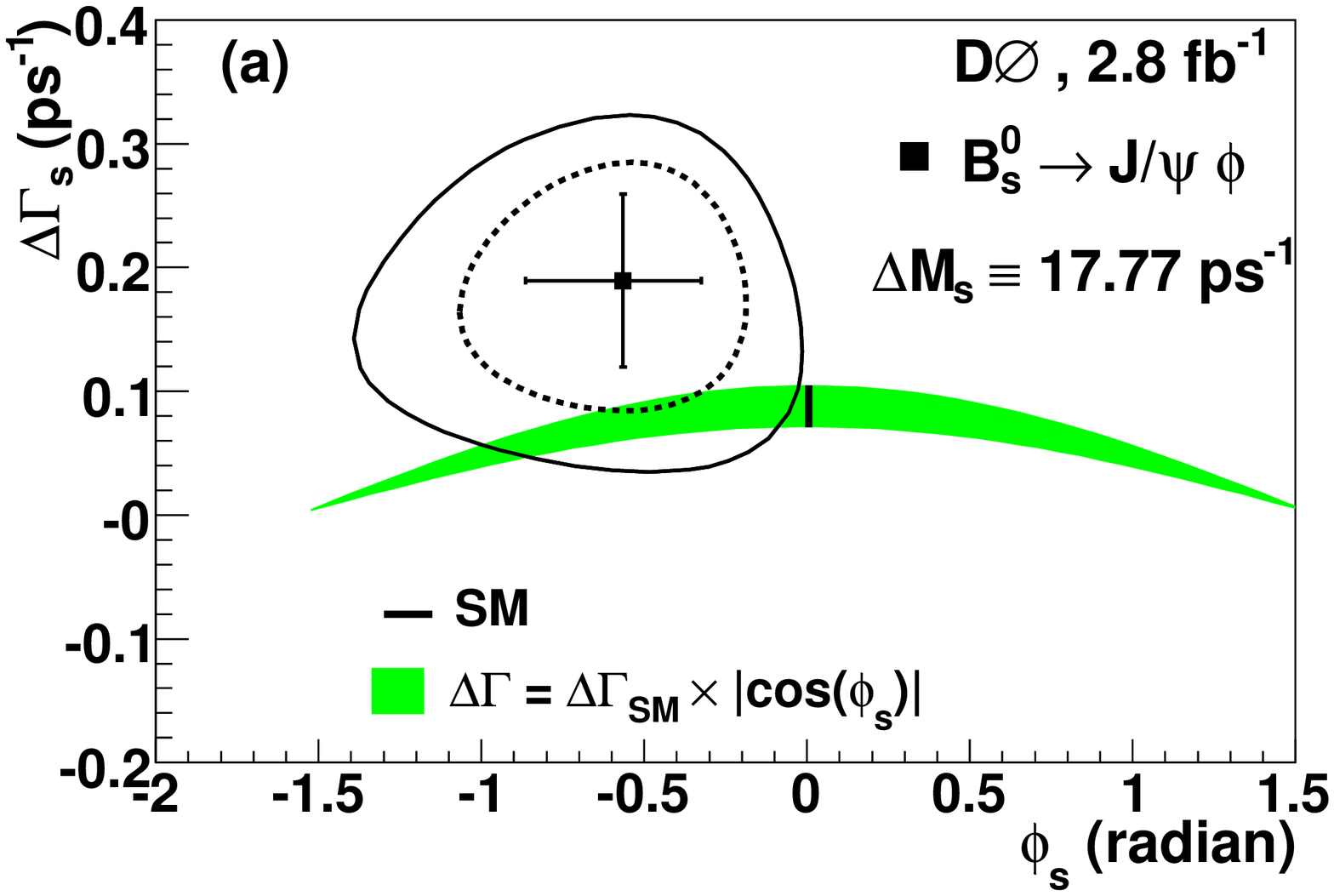,width=7.cm}
\psfig{file=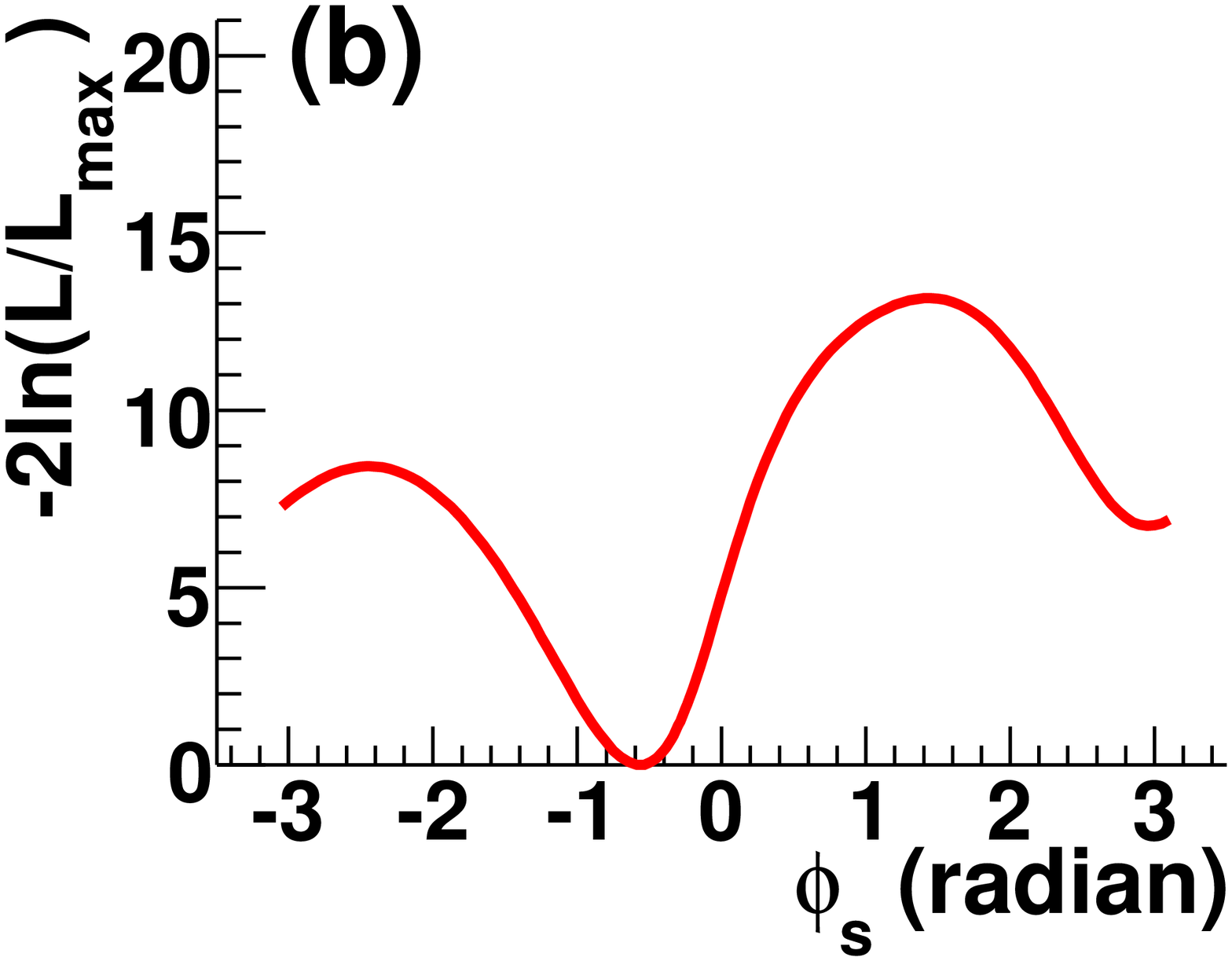,width=4.1cm}
\psfig{file=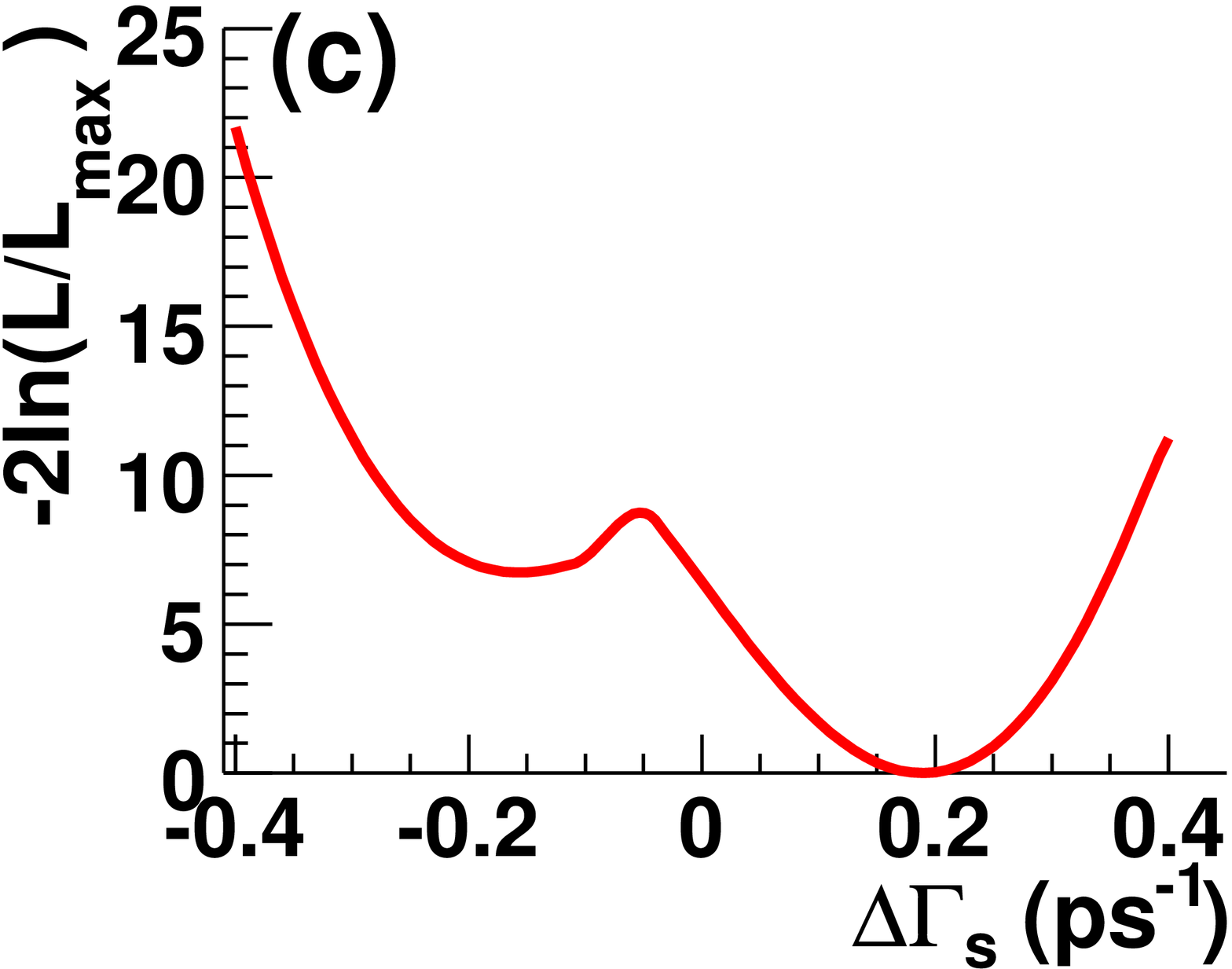,width=4.1cm}
 \end{center}
\caption{
(a) Confidence-level contours in the   
$\Delta\Gamma_s$ - $\phi_s$ plane. The curves  correspond to
expected C.L.= 68.3\% (dashed) and 90\% (solid).
The cross shows  the best
fit point and one-dimensional uncertainties.
Also shown is the SM prediction,  $\phi_s = -2\beta_s = -0.04$,
$\Delta\Gamma_s = 0.088\pm0.017$ ps$^{-1}$~\cite{LN2006}.
(b) Likelihood profile of $\phi_s$,
(c) likelihood profile of $\Delta \Gamma_s$.
}
\label{fig:dgvsphi}
\end{figure}

Results of the fit are presented in Table~\ref{tabmainres}.
The fit yields a likelihood maximum at
$\phi_s = -0.57^{+0.24}_{-0.30}$  and 
$\Delta \Gamma_s = 0.19 \pm 0.07$ ps$^{-1}$, 
where the errors are statistical only.
As a result of the constraints on the phases $\delta_i$,
the second maximum, at 
$\phi_s =2.92^{+0.30}_{-0.24}$, $\Delta \Gamma_s =-0.19 \pm 0.07$ ps$^{-1}$,
is disfavored by a likelihood ratio of 1:29.
Without the constraints on $\delta_i$, $\phi_s$ shifts by only $0.02$ 
for the
$\Delta\Gamma_s>0$ solution.
Confidence level contours in the $\phi_s$  -- $\Delta \Gamma_s$ plane,
and likelihood profiles as a function
of $\phi_s$ and as a function of $\Delta \Gamma_s$ are shown in
 Fig.~\ref{fig:dgvsphi}.
Studies using pseudo-experiments with similar
statistical sensitivity indicate no significant biases
and show that the magnitudes of the statistical uncertainties are
consistent with expectations. The mean value of the statistical uncertainty
in $\phi_s$ from an ensemble generated with the same parameters
as obtained in this analysis is 0.33.
The test finds allowed ranges at the 90\% C.L.
of  $-1.20 <\phi_s < 0.06$ and 
$0.06 <\Delta \Gamma_s  <0.30$  ps$^{-1}$.
To quantify the level of agreement with the SM, we use  pseudo-experiments
with the ``true'' value of the parameter $\phi_s$ set to $-0.04$.
We find the probability of 6.6\% to obtain a fitted value of $\phi_s$
lower than $-0.57$. 


Setting  $\phi_s = -2\beta_s =-0.04$,  as predicted by the SM,
we obtain $\Delta\Gamma_s = 0.14 \pm 0.07$ ps$^{-1}$. This is consistent  
with the theoretical prediction of $0.088\pm0.017$ ps$^{-1}$~\cite{LN2006}.
The results for this fit are shown in the second  column in 
Table~\ref{tabmainres}. 
The non-zero mixing phase is expected to reduce  $\Delta\Gamma_s$
by the factor of $|\cos(\phi_s)|$ compared to its SM value 
 $\Delta\Gamma_s^{SM}$~\cite{DFN2001}.
In the third column of Table~\ref{tabmainres}
we show results of a fit with  $\Delta\Gamma_s$ constrained by this
expected behavior.

The measurement uncertainties are dominated by the limited statistics.
Uncertainty in the acceptance as a function of the transversity angles
is small, the largest effect is on  \ADD.
Effects of the imperfect knowledge of the flavor-tagging purity 
are estimated by
varying the flavor purity parametrization within uncertainties.
The ``interference'' term in the background model accounts for
the collective effect of various physics processes. However, its
presence may be partially due to  detector
acceptance effects. Therefore, we interpret the difference between fits with 
and without this term as a  contribution to the 
systematic uncertainty associated with
 the background model.
The main contributions to 
 system
atic uncertainties for the case of free $\phi_s$ 
are listed in Table \ref{syst}.

\begin{widetext}

\begin{table}[h!tb]
\caption {Sources of  systematic uncertainty in the results  for the case of free $\phi_s$.
}
\begin{tabular}{ccccccccc}
\hline
Source & $\overline \tau_s$ (ps)  & $\Delta \Gamma_s$ (ps$^{-1}$) & $A_{\perp}(0)$  & \ADD\ & \ \ $\phi_s$ \tabularnewline
\hline
\hline

Acceptance                 &$\pm 0.003$  &$\pm 0.003$  &$\pm 0.005$  &$\pm 0.03$  &$\pm 0.005$\tabularnewline
Signal mass model          &$- 0.01$    &$+ 0.006$     &$- 0.003$    &$- 0.001$    &$- 0.006$ \tabularnewline
Flavor purity estimate  &$\pm 0.001$  &$\pm 0.001$  &$\pm 0.001$  &$\pm 0.001$  &$\pm 0.01$ \tabularnewline 
Background model           &$+0.003$     &$+0.02$      &$-0.02$      &$-0.01$      &$+0.02$    \tabularnewline
$\Delta M_s$ input         &$\pm 0.01$  &$\pm 0.001$   &$\pm 0.001$  &$\pm 0.001$  &$+0.06,-0.01$  \tabularnewline
\hline
Total                      &$\pm 0.01$   &$+0.02,-0.01$&$+0.01,-0.02$&$\pm 0.03$   &$+0.07,-0.02$ \tabularnewline

\hline
\hline
\end{tabular}
\label{syst}
\end{table}
\end{widetext}

 In summary, from a fit to the time-dependent angular distribution of 
the flavor-tagged decays \bsdec, 
we have measured the average lifetime 
of the (\bs, \bsbar) system,  $\overline \tau(B^0_s)=1.52 \pm 0.05 \pm 0.01$ ps,  
the  width difference
between  the light  and heavy  $B_s^0$  
eigenstates, 
$\Delta \Gamma_s  = 0.19 \pm 0.07 {\rm (stat)} ^{+0.02}_{-0.01} {\rm (syst)} 
$ ps$^{-1}$, 
and the $CP$-violating phase, $\phi_s = 
-0.57 ^{+0.24}_{-0.30} {\rm (stat)} ^{+0.07}_{-0.02} {\rm (syst)}$.
We also measure the magnitude of the decay amplitudes. In the fits, we set the
oscillation frequency to $\Delta M_s = 17.77$ ps$^{-1}$, 
as measured in Ref.~\cite{dms},
and we impose a Gaussian constraint with a width  of $\pi/5$ to the deviation
of the strong phases from the values
$\delta_1 =-0.46$ and $\delta_2 = 2.92$ of Ref.~\cite{wadelta12}.
The allowed 90\% C.L. intervals of $\Delta \Gamma_s$ and of  $\phi_s$ 
are $0.06 <\Delta \Gamma_s  <0.30$  ps$^{-1}$ and
 $-1.20 <\phi_s < 0.06$.
The SM hypothesis for $\phi_s$ has a $P$-value of 6.6\%.

%
%
%
The results supersede our previous measurements~\cite{prl07}
that were based on the untagged decay \bsdec\ and a smaller data sample.
They are consistent with the recently submitted CDF results~\cite{cdftag}. 

%
We thank U. Nierste for useful discussions.
We thank the staffs at Fermilab and collaborating institutions, 
and acknowledge support from the 
DOE and NSF (USA);
CEA and CNRS/IN2P3 (France);
FASI, Rosatom and RFBR (Russia);
CAPES, CNPq, FAPERJ, FAPESP and FUNDUNESP (Brazil);
DAE and DST (India);
Colciencias (Colombia);
CONACyT (Mexico);
KRF and KOSEF (Korea);
CONICET and UBACyT (Argentina);
FOM (The Netherlands);
PPARC (United Kingdom);
MSMT (Czech Republic);
CRC Program, CFI, NSERC and WestGrid Project (Canada);
BMBF and DFG (Germany);
SFI (Ireland);
The Swedish Research Council (Sweden);
Research Corporation;
Alexander von Humboldt Foundation;
and the Marie Curie Program.
%

This Letter is dedicated to the memory of Andrzej Zieminski.

\end{document}